\documentclass[%
 aps,
 pre,
 superscriptaddress,
 sd,%
 amsmath,amssymb,
reprint,%
showpacs
]{revtex4-1}

\usepackage{graphicx}
\usepackage{dcolumn}
\usepackage{bm}

\usepackage{setspace,color,url}
\newcommand{\mts}{{\mathrm{s}}}
\newcommand{\mtc}{{\mathrm{c}}}
\newcommand{\mta}{{\mathrm{a}}}
\begin{document}

\preprint{Accepted manuscript, not for distribution}

\title{Generic transport coefficients of a confined electrolyte
solution}
\thanks{Accepted manuscript (Phys. Rev. E 90, 052113 (2014)), not for distribution.}
\author{Hiroaki Yoshida}
\email{h-yoshida@mosk.tytlabs.co.jp}
\affiliation{Toyota Central R\&D Labs., Inc., Nagakute, Aichi
 480-1192, Japan}
\affiliation{Elements Strategy Initiative for Catalysts and Batteries
(ESICB), Kyoto University, Kyoto 615-8245, Japan}
\author{Hideyuki Mizuno}
\affiliation{Institut f\"{u}r Materialphysik im Weltraum, Deutsches Zentrum f\"{u}r
Luft- und Raumfahrt (DLR), 51170 K\"{o}ln, Germany}
\author{Tomoyuki Kinjo}
\author{Hitoshi Washizu}
\affiliation{Toyota Central R\&D Labs., Inc., Nagakute, Aichi
 480-1192, Japan}
\affiliation{Elements Strategy Initiative for Catalysts and Batteries
(ESICB), Kyoto University, Kyoto 615-8245, Japan}
\author{Jean-Louis Barrat}%
\affiliation{Laboratory for Interdisciplinary Physics, UMR 5588,
Universit\'e Grenoble 1 and CNRS, 38402 Saint Martin d'H\`eres, France}
\affiliation{Institut Laue--Langevin, 6 rue Jules Horowitz, BP 156, 38042 Grenoble, France}
\date{\today}
%
\begin{abstract}
Physical parameters characterizing electrokinetic transport
in a confined electrolyte solution are reconstructed from the 
generic transport coefficients obtained within  the classical non-equilibrium statistical thermodynamic framework.
The electro-osmotic flow, the diffusio-osmotic flow, the osmotic current,
as well as the pressure-driven Poiseuille-type flow,
the electric conduction, and the ion diffusion, are described by this  set of transport coefficients.
The reconstruction is demonstrated for an aqueous NaCl solution
between two parallel charged surfaces with a nanoscale gap, by using the molecular dynamic (MD) simulations.
A Green--Kubo approach is employed to evaluate the transport
 coefficients in the linear-response regime, 
and the fluxes induced by the pressure, electric, and chemical potential
fields are compared with the results of non-equilibrium MD simulations.
{\color{black}
Using this numerical scheme, the influence of the salt concentration on the
transport coefficients is investigated.}
Anomalous reversal of diffusio-osmotic
 current, as well as that of electro-osmotic flow,
{\color{black} is observed
 at high surface charge densities and high added-salt concentrations.}
\end{abstract}
\maketitle
%
%
%
\section{Introduction\label{sec_intro}}
Controlling and optimizing the
mechanical transports of 
electrolyte solutions in confined geometries
have become increasingly important
in recent remarkable developments of 
electrochemical devices.
Particularly at the scale of nanometer,
novel transport properties
in the vicinity of surfaces
emerge because
of the large surface/volume ratio,
which have potential applications
in various fields,
such as energy conversion~\cite{HBS+2007,SPB+2013},
{\color{black}
water desalination~\cite{BB2010},}
 and fluidic
transistor~\cite{SSH+1999}.
In order to prompt the development
of electrochemical devices 
using the electrokinetic transports,
comprehensive understanding of 
the transport properties is required.

In the context of electrokinetic transports, it is common to focus on 
the mass flow of the solution $Q$ and the electric current ${J}$
induced by the pressure gradient $P$ and the electric field $E$
~\cite{BA2004a,XL2006,LCA+2008}, which are related through the linear relations:
\begin{equation}
\left(\begin{array}{c}
 Q \\ J 
			\end{array}\right) 
=
\left(\begin{array}{ccc}
\widetilde{M}_{11} & \widetilde{M}_{12} \\
\widetilde{M}_{21} & \widetilde{M}_{22} \\
		 \end{array}\right)
\left(\begin{array}{c}
 P \\ E 
			\end{array}\right),
\label{s1-mat0}
\end{equation}
where $\widetilde{M}_{jk}$ denotes the physical transport coefficients.
Note that this equation is valid only in the linear-response regime,
i.e., the system is close to the thermal equilibrium state
such that it responds linearly to the external fields.
We have recently studied
the $Q$ and $J$ of an electrolyte solution in a nano-channel
using  molecular dynamics (MD) simulations, %
showing that a Green--Kubo approach based on the linear-response theory
and the non-equilibrium MD (NEMD) simulation method yield consistent results
in a wide range of the external field strengths~\cite{YMK+2014}.
Along with $Q$ and $J$, however, 
the solute flux is also a very important transport property,
and so is the external field of the solute concentration gradient.
Indeed, an outstanding energy-conversion method utilizing
the diffusio-osmotic current induced by the concentration gradient
has recently been proposed~\cite{SPB+2013}.
In the present study, to realize a systematic investigation into
the electrokinetic transports covering the latter, 
we formulate the transport phenomena in a more general manner
than Eq.~\eqref{s1-mat0}
starting from the classical theory of non-equilibrium thermodynamics,
and apply the scheme to a specific system of a nano-confined electrolyte solution.

\section{Formulation of the transport coefficients\label{sec_intro}}
We consider an electrolyte solution
consisting of three components,
namely, a solvent, a cation, and an anion.
Then the system responses to the external forces
due to the chemical potential gradients
of each component, in the linear-response regime, are written in the
following form:
\begin{equation}
\left(\begin{array}{c}
{N_\mts} \\ {N_\mtc} \\ {N_\mta}
			\end{array}\right) 
=
\left(\begin{array}{ccc}
M_{\mts\mts} & M_{\mts\mtc} & M_{\mts\mta}\\
M_{\mtc\mts} & M_{\mtc\mtc} & M_{\mtc\mta}\\
M_{\mta\mts} & M_{\mta\mtc} & M_{\mta\mta}
		 \end{array}\right)
\left(\begin{array}{c}
 F_\mts \\ F_\mtc \\ F_\mta
			\end{array}\right),
\label{s1-mat}
\end{equation}
where $N_{\alpha}$ with $\alpha=\mts,\mtc,$ and $\mta$
denotes the molar fluxes of the solvent, cation, and anion, respectively,
and $F_{\alpha}$ is the force per mole representing the chemical potential
gradient. 
The transport coefficients
are evaluated from the
time-correlation function of
the fluctuated fluxes $N_{\alpha}$ at thermal equilibrium,
through the Green--Kubo formula derived from  linear-response theory~\cite{BB1994,MDJ+2003,HM2006,BB2013}:
\begin{equation}
M_{\alpha \beta}
=\dfrac{V}{k_{\mathrm{B}}T}\int_0^{\infty}
\langle N_{\alpha}(t)N_{\beta}(0)\rangle\mathrm{d}t,
\label{s1-g-k}
\end{equation}
where $V$ is the system volume, $T$ is the temperature, and
$k_{\mathrm{B}}$ is the Boltzmann constant.
In a  system with a microscopic dynamics that is invariant under time reversal,  the correlation $\langle N_{\alpha}(t)N_{\beta}(0)\rangle$
is statistically identical to $\langle
N_{\beta}(t)N_{\alpha}(0)\rangle$.
Hence, the relation $M_{\alpha\beta}=M_{\beta\alpha}$ holds,
which is known as Onsager's reciprocal relation~\cite{O1931A,*O1931B,DM1962}.
The matrix in Eq.~\eqref{s1-mat} is  thus symmetric,
and the number of independent coefficients in Eq.~\eqref{s1-mat} is six.

Once the six coefficients $M_{\alpha\beta}$ have been estimated,
all the transports phenomena in the electrolyte solution in response to
a weak external force are covered.
In experiments, however, one usually measures
fluxes that are different from the $N_{\alpha}$ set.
Note that since the degree of freedom of the 
three component system is three, 
there should be three fluxes characterizing the
transport phenomena, which are expressed in terms of linear
combination of $N_{\alpha}$.
A set of fluxes that
is commonly used in experiments consists of the mass flow $Q$, the electric current $J$, and
the solute flux $G$.
These fluxes are
expressed in terms of the component fluxes $N_{\alpha}$ in the following
form:
\begin{equation}
\left(\begin{array}{c}
 Q \\ J \\ G
			\end{array}\right) 
=
\left(\begin{array}{ccc}
m_\mts & m_\mtc & m_\mta\\
0 & z_{\mtc}eN_{\mathrm{A}} & -z_{\mta}eN_{\mathrm{A}}\\
0 & 1 & 1
		 \end{array}\right)
\left(\begin{array}{c}
{{N}_\mts} \\ {{N}_\mtc} \\ {{N}_\mta}
			\end{array}\right),
\label{s1-flux}
\end{equation}
where $m_{\alpha}$ is the mass per mole, $e$ is the unit charge,
$z_{\alpha}$ is the valence, and $N_{\mathrm{A}}$ is the Avogadro number.
The corresponding
external fields are then
the pressure gradient $P$, the electric field $E$,
and the  gradient of the solute chemical potential, denoted by  $H$.
The relation among these external fields and $F_{\alpha}$ is
\begin{equation}
\left(\begin{array}{c}
{F_\mts} \\ {F_\mtc} \\ {F_\mta}
			\end{array}\right) 
=
\left(\begin{array}{ccc}
m_\mts & 0 & 0 \\
m_\mtc & z_{\mtc}eN_{\mathrm{A}} & 1 \\
m_\mta & -z_{\mta}eN_{\mathrm{A}} & 1
		 \end{array}\right)
\left(\begin{array}{c}
 P \\ E \\ H
			\end{array}\right).
\label{s1-force}
\end{equation}
The fluxes 
$Q$, $J$, and $G$ 
that linearly respond to the external fields
$P$, $E$, and $H$
are
then written in the following form:
\begin{equation}
\left(\begin{array}{c}
 Q \\ J \\ G
			\end{array}\right) 
=
\left(\begin{array}{ccc}
\widetilde{M}_{11} & \widetilde{M}_{12} & \widetilde{M}_{13}\\
\widetilde{M}_{21} & \widetilde{M}_{22} & \widetilde{M}_{23}\\
\widetilde{M}_{31} & \widetilde{M}_{32} & \widetilde{M}_{33}
		 \end{array}\right)
\left(\begin{array}{c}
 P \\ E \\ H
			\end{array}\right),
\label{s1-mat2}
\end{equation}
where 
$\widetilde{M}=SMS^{\mathrm{T}}$
with $M$ and $S$ 
being the matrices in Eqs.~\eqref{s1-mat} and \eqref{s1-flux},
respectively.
Note that Onsager's reciprocal relations
are  preserved in this transformation
($\widetilde{M}_{jk}=\widetilde{M}_{kj}$.)
In addition to the components of $2\times 2$ matrix appearing in
Eq.~\eqref{s1-mat0},
the physical parameters in relation to $G$ and $H$ are included in $\widetilde{M}$;
for instance, $\widetilde{M}_{23}$ corresponds to
the diffusio-osmotic current,
and $\widetilde{M}_{32}$ to the electro-osmotic diffusion.
Although one might choose different
set of fluxes than $Q$, $J$, and $G$, depending on the 
experimental setup, once the generic transport coefficients
in Eq.~\eqref{s1-mat} are evaluated, all the 
physical parameters characterizing the transport phenomena are
obtained straightforwardly,  the mapping $S$ above serving  as an
example. For instance, one can easily, by an appropriate transformation,
define  the appropriate transport coefficients   in a situation
in which one of the ionic currents is blocked while the second one is nonzero.

\begin{figure}[t]
\begin{center}
\centering
\includegraphics[scale=0.8]{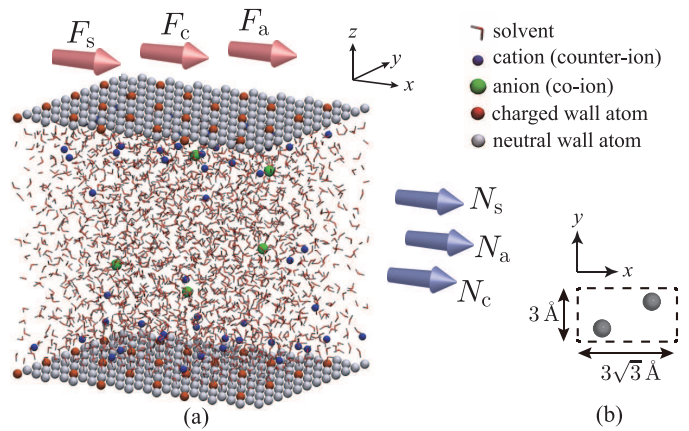} 
\caption{
\color{black}
(a) An electrolyte solution between two charged surfaces.
(b) A unit cell of the equilateral triangular lattice of the wall atoms in the $x$-$y$ plane.
}
\label{fig01}
\end{center}
\end{figure}  
%

%
%
%
\section{Application to aqueous $\textrm{NaCl}$ solution\label{sec_application}}

%
\subsection{Molecular dynamics simulation\label{sec_md}}
We apply the formulation described in the previous section
to a system consisting of an aqueous NaCl solution confined between
two parallel charged surfaces {\color{black}as shown in Fig.~\ref{fig01}}.
The molecular dynamics simulation method is employed,
because it allows an efficient, detailed analysis 
of the microscopic phenomena at the atomic  scale~\cite{RP2013}.
Each wall consists of  a 
two-dimensional  triangular lattice
of a model atom, 
with a lattice spacing  $a_0=3$\,{\AA}. A charged, periodic
superstructure with a periodicity $\ell_c=3a_0$ 
{\color{black} or $5a_0$}
 is superimposed onto this
triangular lattice, so that one atom out of 9 {\color{black}or 25} is 
 negatively charged with one unit charge $-e$. The 
resulting surface  charge density is $\sigma=0.228$\,C/m$^2$
in the case of $\ell_c=3a_0$, 
{\color{black} and $\sigma=0.082$\,C/m$^2$ in the case of
$\ell_c=5a_0$.}
The numbers of Na$^+$ and  Cl$^-$ ions in the electrolyte solution
are denoted by $n^{\mathrm{Na}}$ and $n^{\mathrm{Cl}}$, respectively.
Then the relation $n^{\mathrm{Na}}=n^{\mathrm{Cl}}+n^{\mathrm{C}}$
holds
because of electrical neutrality, where  $n^{\mathrm{C}}$ is the number of charged wall atoms.
The  interactions between water molecules are
described by the extended simple point charge (SPC/E) model,
and those between ions are
described by a sum of electrostatic and Lennard-Jones (LJ) potentials,
with parameters taken from Ref.~\cite{SD1994}.
The Lorentz--Berthelot mixing rule~\cite{AT1989}
is employed for the LJ parameters for water-ion and Na-Cl pairs.
For the interaction between a wall atom and a water molecule,
a model mimicking a hydrophilic surface at the level
of homogeneously distributed hydrogen bond sites is used,
{\color{black}
where the potential is designed to have orientation dependence
reflecting the trend of hydrogen bonds
\cite{PG2008,YMK+2014}.}
The parallel code LAMMPS~\cite{LAMMPS}
is used to implement the MD simulation.
The number of particles and the volume $V$ are kept constant,
and the Nos\'e--Hoover thermostat is used 
to maintain the temperature at $T=300$\,K  (NVT ensemble).
Further details of the computational procedure are
 described in Ref.~\cite{YMK+2014}.

\begin{table}[t]
\centering
\caption{Transport coefficients in unit $10^{-9}$mol$^2$/Jms; values in parentheses are
 standard errors for ten simulation runs.}
\label{table01}
\vspace*{2mm}
\renewcommand{\arraystretch}{1.0}
\begin{tabular}{cccc}
\hline
\hline
flux\textbackslash force & $F_\mts$ & $F_\mtc$ & $F_\mta$ \\
\hline
$N_\mts$ & $M_{\mts\mts}$ & $M_{\mts\mtc}$ & $M_{\mts\mta}$\\ 
    &$3754$ ($74$) & $14.51$ ($0.94$) & $10.69$ ($0.40$)\\[3pt]
$N_\mtc$ & $M_{\mtc\mts}$ & $M_{\mtc\mtc}$ & $M_{\mtc\mta}$\\ 
    & $14.05$ ($0.99$)& $0.2526$ ($0.0063$)& $0.04447$ ($0.0064$)\\ [3pt]
$N_\mta$ & $M_{\mta\mts}$ & $M_{\mta\mtc}$ & $M_{\mta\mta}$\\ 
    & $10.88$ ($0.72$)& $0.04317$ ($0.0069$)& $0.1390$ ($0.0040$)\\ 
\hline
\hline
\end{tabular}
\end{table}
%

%
\subsection{Transformation of the generic transport coefficients\label{sec_tc}}
{\color{black}
We first demonstrate
the transformation from the generic transport coefficients $M_{\alpha\beta}$ to
the physical parameters $\widetilde{M}_{jk}$.}
The transport coefficients are evaluated
using the Green--Kubo formula \eqref{s1-g-k},
for the system
of the walls with lateral dimensions of $4.68$\,nm\,$\times 3.6$\,nm
{\color{black}
($9\times 12$ unit cells, see Fig.~\ref{fig01})
and $\sigma=0.228$\,C/m$^2$ ($\ell_c=3a_0$),}
confining $2235$ water molecules,
$53$ Na$^+$ ions, and $5$ Cl$^-$ ions.
The distance between wall atoms 
determined using the method described in Ref.~\cite{YMK+2014}
is $4.12$\,nm,
and the resulting nominal concentrations of Na$^+$ and
Cl$^-$ are $1.27$\,M and $0.12$\,M, respectively.
Because of poor statistics in the equilibrium simulations,
an extremely long simulation run is required to
obtain  smooth time-correlation functions in Eq.~\eqref{s1-g-k}.
To circumvent this difficulty,
we carry out 
ten MD simulations with different initial configurations,
each of which runs for $5$\,ns,
and time-integrate the averaged correlation functions.

Table~\ref{table01} lists the values of $M_{\alpha\beta}$,
along with the standard errors for ten simulation runs.
Onsager's reciprocal relations are satisfied reasonably well,
which shows the good accuracy of the numerical simulations.
Although the bare components of $M_{\alpha\beta}$ 
are different from the familiar physical parameters,
they give some  interesting indications.
The higher mobility of the excess counter-ions (cations) compensating the 
negative surfaces charges is indicated by $M_{\mtc\mtc}>M_{\mta\mta}$.
Regarding cross effects, the clear difference   $M_{\mts\mtc}>M_{\mts\mta}$ implies
occurrence of the electro-osmotic flow,
because the values of $M_{\mts\mtc}$ and $M_{\mts\mta}$ represent the 
intensity of the solvent flow induced by the force acting on ions.

\begin{figure}[t]
\begin{center}
\centering
\includegraphics[scale=1.3]{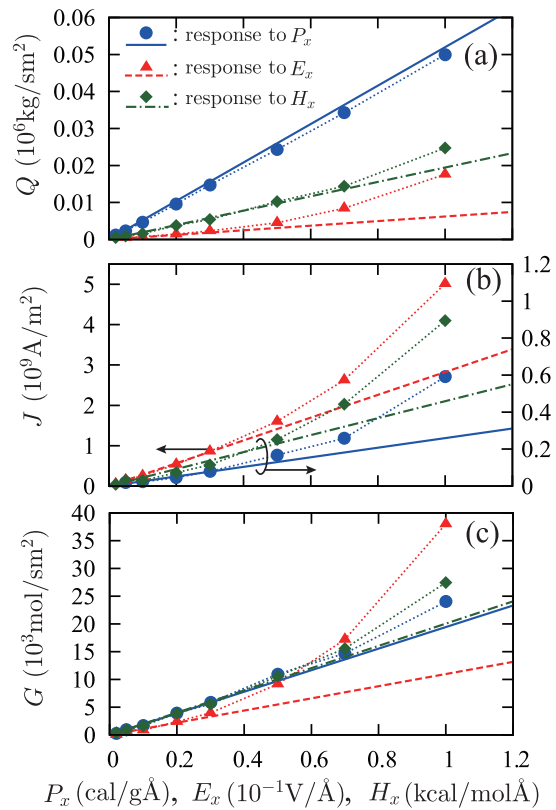} 
\vspace{-1mm}
\caption{
(a) Mass flow density, (b) current density,
and (c) diffusion flux density, induced by 
the pressure gradient, the electric field, and the chemical potential gradient.
The linear lines
indicate Eq.~\eqref{s1-mat2} based on the generic transport coefficients,
and the symbols indicate the results of the NEMD simulations with the
explicit external fields.
}
\label{fig02}
\end{center}
\end{figure}  

The generic transport coefficients
are readily transformed into the physical parameters 
that are relevant in the situation
under which the measurements are performed.
Since the most common set of observable quantities
for the confined electrolyte solution is $Q$, $J$, and $G$ introduced above,
we carry out the NEMD simulations
and numerically obtain these fluxes
to ensure that the transformation of the transport coefficients works,
and to examine the limit of the linear-response assumption.
The external forces exerted on $i$th particle in the $x$-direction are 
$f_i^{\mathrm{ext}}=m_iP_x$
where $P_x$ is the mass acceleration simulating the pressure
gradient, $f_i^{\mathrm{ext}}=q_iE_x$
where $E_x$ is the electric field, and
$f_i^{\mathrm{ext}}=\psi^{\mathrm{ion}}_iH_x$ where
$H_x$ is the force representing the chemical potential gradient of the
solute; $m_i$ and $q_i$ are the mass and charge of $i$th particle,
and $\psi^{\mathrm{ion}}_i$ is an index of which the value is unity for
ions and zero for solvent particles.
After an equilibriation for $1$\,ns, a production run
for $4$\,ns is performed at a set of external fields specified,
to obtain average values of the mass flow density
$Q=(1/V)\sum_im_i\dot{x}_i$,
the electric current density $J=(1/V)\sum_iq_i\dot{x}_i$, and
the ion-flux density $G=(1/V)\sum_i\psi_i^{\mathrm{ion}}\dot{x}_i$,
which are plotted in Fig.~\ref{fig02}.
All the fluxes 
approach asymptotically as $P_x, E_x, H_x \to 0$,
showing that the numerical values of the coefficients in
Table~\ref{table01} correctly reproduce
the physical parameters of $\widetilde{M}$ in Eq.~\eqref{s1-mat2}.
The linear-response assumption is
valid in the range 
$P_x\le 0.2$\,cal/g{\AA}, $E_x\le 0.02$\,V/{\AA}, and $H_x\le 0.2$\,kcal/mol{\AA}.
Note that these critical values are 
extremely large compared with
the field strength attainable in laboratories;
for instance
$P_x=0.2$\,cal/g{\AA} along a distance of $1$\,$\mu$m results in 
a pressure difference of $10^{5}$ atm. 
Figure~\ref{fig02} indicates the linear-response assumption to be valid
in quite wide range of the field strengths in the system considered herein.

\begin{figure}[t]
\begin{center}
\centering
\includegraphics[scale=1.4]{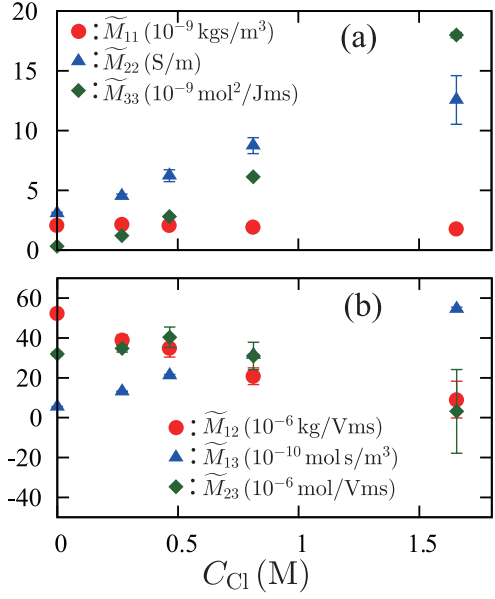} 
\caption{
\color{black}
Physical transport coefficients $\widetilde{M}_{jk}$
as functions of the concentration of Cl$^{-}$ 
at the midpoint of the channel,
in the case of $\sigma=0.082$\,C/m$^2$. (a) Diagonal components,
and (b) off-diagonal components. The error bar indicates 
the standard errors for ten simulation runs.
}
\label{fig03}
\end{center}
\end{figure}  
\begin{figure}[t]
\begin{center}
\centering
\includegraphics[scale=1.4]{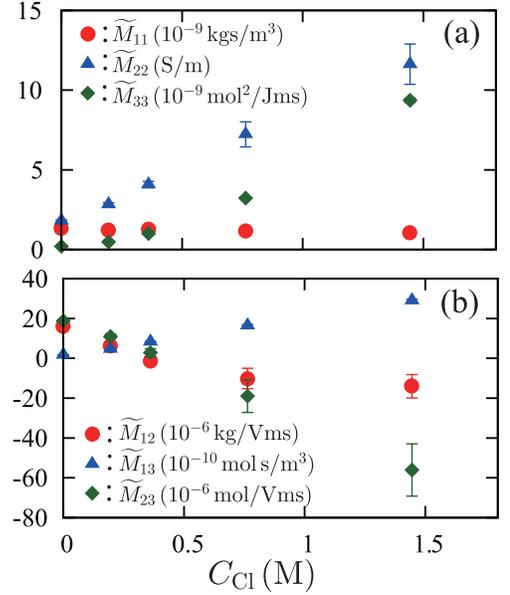} 
\caption{
Physical transport coefficients $\widetilde{M}_{jk}$
as functions of the concentration of Cl$^{-}$,
in the case of $\sigma=0.228$\,C/m$^2$. 
See the caption of Fig.~\ref{fig03}.
}
\label{fig04}
\end{center}
\end{figure}  
%

%
\subsection{Influence of salt concentration and flow reversal\label{sec_sc}}
At relatively high surface charge densities,
the reversal of the  electro-osmotic flow, i.e.,
the negative response of $Q$ to $E_x$, can occur
as first demonstrated by Qiao and Aluru~\cite{QA2004}.
Recently we have shown the occurrence of
the reversal of the electro-osmotic flow, as well as its reciprocal streaming current,
in the linear-response regime~\cite{YMK+2014}.
In addition to the surface charge density,
importance of the concentration of the added salt on the
transport properties was also
implied in Ref.~\cite{YMK+2014}.
Here, we examine systematically the influence of the added salt.
Specifically, maintaining the surface charge density at 
{\color{black}
$\sigma=0.082$ and 
$0.228$\,C/m$^2$,}
the concentration of the added salt is 
varied by controlling the number of excess pairs of Na$^+$ and Cl$^-$.

\begin{figure}[t]
\begin{center}
\centering
\includegraphics[scale=1.3]{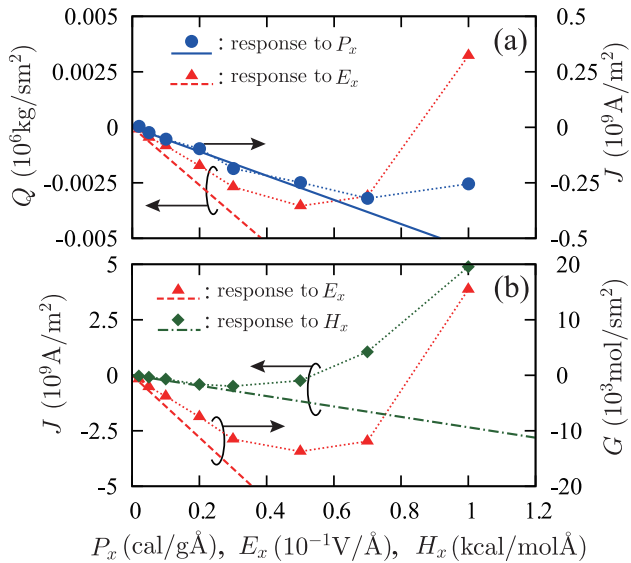} 
\caption{
(a) Inverted electro-osmotic flow and the reciprocal
streaming current,
and (b) inverted electro-osmosis and the reciprocal diffusio-osmotic current,
 observed at $\sigma=0.228$\,C/m$^2$ and $C_{\mathrm{Cl}}=1.44$\,M.
See the caption of Fig.~\ref{fig02}.
}
\label{fig05}
\end{center}
\end{figure}

{\color{black}
In Figs.~\ref{fig03} and \ref{fig04},}
the components of the matrix $\widetilde{M}$ are plotted
as functions of the bulk concentration of Cl$^-$, denoted by
$C_{\mathrm{Cl}}$, 
where $C_{\mathrm{Cl}}$ is measured at the midpoint of the gap.
{\color{black}
In these parameter ranges,
the nominal concentration of Na$^+$ ranges
from $0.41$ to $1.57$\,M in Fig.~\ref{fig03},
and it ranges from $1.15$ to $2.30$\,M in Fig.~\ref{fig04}.
The weak dependence of $\widetilde{M}_{11}$ on the salt concentration
indicates that the rate of the pressure-driven Poiseuille-type flow
is not influenced significantly, implying the
weak variation of the kinetic viscosity of the electrolyte solution
in this parameter range.
The coefficients $\widetilde{M}_{22}$ and $\widetilde{M}_{33}$,
corresponding respectively to the electrical conductivity and
the salt diffusivity, increase as the salt concentration
due to the increase of the carrier, and so does $\widetilde{M}_{13}$.
In contrast, $\widetilde{M}_{12}$ 
and $\widetilde{M}_{23}$ decrease
as the salt concentration.
Particularly in the case of $\sigma=0.228$\,C/m$^2$ (Fig.~\ref{fig04}(b)),}
the values of $\widetilde{M}_{12}$ (and $\widetilde{M}_{21}$),
and $\widetilde{M}_{23}$ (and $\widetilde{M}_{32}$) become
negative for high concentrations ($C_{\mathrm{Cl}}>0.4$\,M),
meaning that, in addition to the
the electro-osmotic flow and the streaming current,
 the diffusio-osmotic current (response of $J$ to $H_x$)
and its reciprocal electro-osmotic diffusion (response of $G$ to $E_x$)
are anomalously reversed.
Here, we note that the matrix $\widetilde{M}$
maintains the positive definiteness for all cases.
The fluxes obtained via Eq.~\eqref{s1-mat2}, corresponding to 
$\widetilde{M}_{12}$, $\widetilde{M}_{21}$, $\widetilde{M}_{23}$,
and $\widetilde{M}_{32}$, at $C_{\mathrm{Cl}}=1.44$\,M are shown in Fig.~\ref{fig05},
along with the results of the NEMD simulations.
Although the nonlinear effect at  extremely large external fields
changes the direction of the fluxes,
the results of the NEMD simulations converge 
to the values predicted by the transport coefficients
in the linear-response regime, which confirms the occurrence
of the reversed responses.

\begin{figure}[t]
\begin{center}
\centering
\includegraphics[scale=1.3]{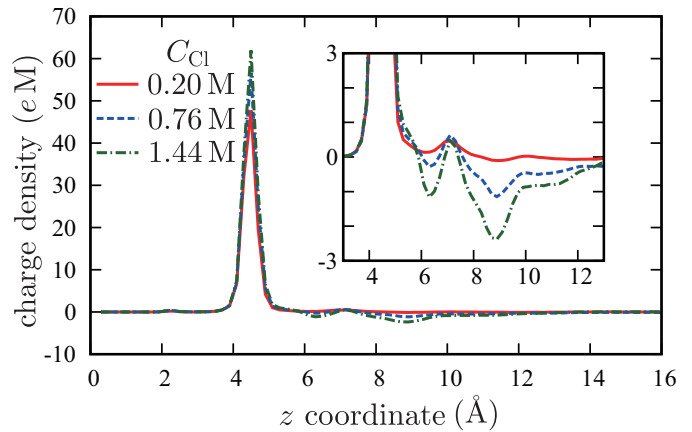} 
\caption{\color{black}
Charge density distributions in the $z$-direction,
for the case of $\sigma=0.228$\,C/m$^2$.
The origin of the coordinate is at the position of the wall atoms.
The inset is the same data on a smaller scale.
}
\label{fig06}
\end{center}
\end{figure}

It was shown in Refs.~\cite{QA2004,YMK+2014}
that the strong binding or the counter-ion condensation
at high surface charge densities makes 
co-ions gather 
in the region where the solution can move, which results
in the reversed electro-osmotic flow.
{\color{black}
We show in Fig.~\ref{fig06}
the distribution of the net charge across the channel width
for several values of the salt concentration
in the case of $\sigma=0.228$\,C/m$^2$.
At high salt concentrations, a negatively charged region is observed 
around $z=9$\,{\AA}.
}
The existence of the negative net charge in the mobile region
also explains the reversed diffusio-osmotic current and 
electro-osmotic diffusion. Although 
the average concentration of  the counter-ions compensating the surface charge is large,
most of them condense at the charged surface and 
do not respond to the diffusion force or to the electric field.
As a result, the number of co-ions accumulating in the mobile region
exceeds that of counter-ions there,
which causes diffusio-osmotic current and electro-osmotic diffusion
in the direction opposite to the usual ones.

%

%
\section{Conclusion\label{sec_conclusion}}
We have described
the transformation
of the generic  transport coefficients $M_{\alpha\beta}$
for a confined electrolyte solution
into the physical transport coefficients 
$\widetilde{M}_{jk}$,
which preserves Onsager's reciprocal relation.
Applicability of the transformation
has been demonstrated
by using 
the equilibrium and NEMD simulations for the
system of an aqueous NaCl solution
confined in a charged nano-channel. The generic coefficients are
obtainable in the standard framework of equilibrium molecular dynamics
and Green--Kubo formula, while the physical ones are more naturally
obtained using the NEMD simulations in the limit of small external fields.
{\color{black} The influence of the salt concentration
on the transport coefficients has been investigated
at two values of the surface charge densities. Our results are expected to be generic, and 
 provide
important information for the design of electrochemical devices using nano-porous media.
}
Furthermore, anomalous reversal of the
diffusio-osmotic current,
as well as the reversal of the electro-osmotic flow,
at high surface charge density and high
concentration of added salt, has been
shown to occur in the linear-response regime.

The usefulness of the transformation
would be more pronounced for 
complex systems with larger number of chemical components
in the solution, because 
any physical parameters
of interest in an experimental setup are immediately 
obtained via the transformation from the 
generic transport coefficients,
which are evaluated in a straightforward manner 
simply using the fluxes of each component
as shown in Eq.~\eqref{s1-g-k}.
{\color{black}
Our future work thus includes application of the
presented scheme 
to different chemical components and to wider parameter ranges,
possibly using the coarse-grained molecular simulation (e.g., Ref.~\cite{KYW2014}), in systems 
for which the all atom molecular dynamics simulation is not feasible.}


\begin{acknowledgments}
The authors thank S. Iwai for computer assistance.
H.~Y., T.~K., and H.~W. are supported by MEXT program ``Elements
Strategy Initiative to Form Core Research Center'' (since 2012).  (MEXT
stands for Ministry of Education, Culture, Sports, Science, and
Technology, Japan.)
H.~M. acknowledges support from DAAD (German Academic Exchange Service.)
J.-L.~B. is supported by the Institut Universitaire de France, and acknowledges useful discussions with L. Bocquet and E. Charlaix.
\end{acknowledgments}


%

\end{document}